\documentclass[twocolumn,showpacs,preprintnumbers,amsmath,amssymb]{revtex4}

\usepackage{graphicx}
\usepackage{dcolumn}
\usepackage{bm}
\usepackage{amsmath}
\usepackage{amsfonts}
\usepackage{color}
\input{Qcircuit}
\begin{document}

\preprint{APS/123-QED}

\title{Application of Darboux Transformation to solve Multisoliton Solution on Non-linear Schr\"odinger Equation}

\author{Agung Trisetyarso}
\affiliation{Department of Applied Physics and Physico-Informatics, Keio University, Yagami Campus\\
3-14-1 Hiyoshi, Kohoku-ku, Yokohama-shi, Kanagawa-ken 223-8522, Japan} 

\date{\today}

\begin{abstract}

Darboux transformation is one of the methods used in solving nonlinear evolution equation. Basically, the Darboux transformation is a linear algebra formulation of the solutions of the Zakharov-Shabat system of equations associated with the nonlinear evolution equation. In this work, the evolution of monochromatic electromagnetic wave in a nonlinear-dispersive optical medium is considered. Using the Darboux transformation, explicit multisoliton solutions (one to three soliton solutions) are obtained from a trivial initial solution.


\end{abstract}

\maketitle

\section[Introduction]{\label{sec:level1}Introduction}

In 1882, Gaston Darboux introduced a method to solve Sturm-Liouville differential equation, which is called Darboux transformation afterwards\cite{Matveev01}. Remarkably, this transformation can solve not only Sturm-Liouville differential equation (another form of time-independent Schr\"odinger equation), but also for the other forms of linear and non-linear differential equations, such as Korteweig-de Vries, Kadomtsev-Petviashvilli, Sine-Gordon, and Non-linear Schr\"odinger equations. Following exposition explains about Darboux transformation and its generalized form called Crum theorem on Sturm-Liouville differential equation.   

\section[Darboux Transformation and Crum Theorem]{\label{sec:level1}Darboux Transformation and Crum Theorem}

Consider following Sturm-Liouville differential equation
\begin{equation}
\label{SLE}
-\psi_{xx}+u\psi=\lambda \psi
\end{equation}
\noindent
where $u$ is function of $x$ and $\lambda$ is a constant. In the Schr\"odinger equation, $u$ represent a potential. Darboux transformation is defined as follow\cite{Matveev01}
\begin{eqnarray}
\label{SE}
\psi[1]=\left(\frac{d}{dx}-\sigma_{1}\right)\psi=\psi_{x}-\frac{\psi_{1x}}{\psi_{1}}\psi \nonumber \\
=\frac{\psi_{x}\psi_{1}-\psi_{1x}\psi}{\psi_{1}} = \frac{W\left(\psi_{1}, \psi\right)}{W\left(\psi_{1}\right)}
\end{eqnarray}
\noindent
$W\left(\psi, \psi[1]\right)$ expresses Wronskian determinant as below
\begin{equation}
\label{DWr}
W(\psi_{1}, \psi_{1}, ... \psi_{N}) = \left| \begin{array}{cccc}
\psi_{1} & \psi_{2} & ... & \psi_{N} \\
\psi_{1}^{(1)} & \psi_{2}^{(1)} & ... & \psi_{N}^{(1)} \\
... & ... & ... & ... \\
\psi_{1}^{(n-1)} & \psi_{2}^{(n-1)} & ... & \psi_{N}^{(n-1)} \end{array} \right|.
\end{equation}
\noindent
where $\psi^{(n)}$ means $\psi$ derivative for $n$-times and $\psi_{1}$ is the solution of $\psi$ for $\lambda = \lambda_{1}$. If $\psi$ is a solution, $\psi[1]$ is the solution for the following Sturm-Liouville differential equation then
\begin{equation}
\label{DTSL}
-\psi_{xx}[1]+u[1]\psi[1]=\lambda \psi[1]
\end{equation}
\noindent
where $u[1]$ is a new function of $u$ which has been transformed. It will be shown below that Darboux transformation acting on $\psi[1]$ influences the potential $u$, if $\psi[1]$ is invariant over Eq. (\ref{DTSL}). 

From Eq.(\ref{SE}), one can obtain following relation
\begin{equation}
\label{SED}
-\psi_{xx}[1]=-\psi_{xxx}+\sigma_{1xx}\psi+2\sigma_{1x}\psi_{x}+\sigma_{1}\psi_{xx}
\end{equation}
\noindent
Substitution of Eq.(\ref{SE}) and Eq.(\ref{SED}) into Eq.(\ref{DTSL}) results
\begin{eqnarray}
\label{sbt}
-\psi_{xxx}+\sigma_{1xx}\psi+2\sigma_{1x}\psi_{x}+\sigma_{1}\psi_{xx}+u[1]\left(\psi_{x}-\sigma_{1}\psi\right) \nonumber
 \\=\lambda \left(\psi_{x}-\sigma_{1}\psi\right)
\end{eqnarray}
\noindent 
Using Eq.(\ref{SLE}) to substitute $\psi_{xx}$ results
\begin{eqnarray}
\label{sbt2}
\left(u[1]-u+2\sigma_{1x}\right)\psi_{x}+ \nonumber
\\ \left(-u_{x}+\sigma_{1xx}+\sigma_{1}u-\sigma_{1}u[1]\right)\psi=0
\end{eqnarray}
\noindent
It is clear from Eq.(\ref{sbt2}), the following relations can be obtained
\begin{eqnarray}
\label{s2}
u[1]=u-2\sigma_{1x}
\\ \sigma_{1xx}-u_{x}+2\sigma_{1}\sigma_{1x}=0
\end{eqnarray}
\noindent
The Eq.(\ref{SED}) shows that function $u$ is also under the transformation due to $\psi$ in Eq.(\ref{SLE}) under Darboux transformation. In other words, Sturm-Liouville equation in Eq.(\ref{SLE}) is covariant under Darboux transformation action
\begin{equation*}
\label{s2a}
\psi \rightarrow \psi[1] ~\text{and}~u \rightarrow u[1]
\end{equation*} 
\indent
Interestingly, Darboux transformation can be recursively applied to Sturm-Liouville equation solution and consequently the potential is under transformation to ensure that the solution belongs to the equation. 

\section[Discussion]{\label{sec:level1a}Discussion}

The implementations of Darboux transformations have been shown in the cases of graphene \cite{trisetyarso2012dirac} and cavity quantum electrodynamics \cite{trisetyarso2010correlation}  \cite{trisetyarso2011erratum} and how this would contribute into the problem of group theory is still explored \cite{trisetyarso2008degeneracy}.  The scheme can be related to the problem of perturbation theory as well \cite{trisetyarso2013perturbation}. This may contribute into the problem of measurement-based quantum computation  \cite{trisetyarso2009measurement} \cite{trisetyarso2009resources} \cite{trisetyarso2011theoretical}.


\begin{thebibliography}{9}
\expandafter\ifx\csname natexlab\endcsname\relax\def\natexlab#1{#1}\fi
\expandafter\ifx\csname bibnamefont\endcsname\relax
  \def\bibnamefont#1{#1}\fi
\expandafter\ifx\csname bibfnamefont\endcsname\relax
  \def\bibfnamefont#1{#1}\fi
\expandafter\ifx\csname citenamefont\endcsname\relax
  \def\citenamefont#1{#1}\fi
\expandafter\ifx\csname url\endcsname\relax
  \def\url#1{\texttt{#1}}\fi
\expandafter\ifx\csname urlprefix\endcsname\relax\def\urlprefix{URL }\fi
\providecommand{\bibinfo}[2]{#2}
\providecommand{\eprint}[2][]{\url{#2}}

\bibitem[{\citenamefont{V.~Matveev}(1990)}]{Matveev01}
\bibinfo{author}{\bibfnamefont{M.~S.} \bibnamefont{V.~Matveev}},
  \emph{\bibinfo{title}{Darboux Transformations and Solitons}}, Springer Series
  in Nonlinear Dynamics (\bibinfo{publisher}{Springer}, \bibinfo{year}{1990}).

\bibitem[{\citenamefont{Trisetyarso}(2012)}]{trisetyarso2012dirac}
\bibinfo{author}{\bibfnamefont{A.}~\bibnamefont{Trisetyarso}},
  \bibinfo{journal}{Quantum Information \& Computation}
  \textbf{\bibinfo{volume}{12}}, \bibinfo{pages}{989} (\bibinfo{year}{2012}).

\bibitem[{\citenamefont{Trisetyarso}(2010)}]{trisetyarso2010correlation}
\bibinfo{author}{\bibfnamefont{A.}~\bibnamefont{Trisetyarso}},
  \bibinfo{journal}{Journal of Mathematical Physics}
  \textbf{\bibinfo{volume}{51}}, \bibinfo{pages}{072103}
  (\bibinfo{year}{2010}).

\bibitem[{\citenamefont{Trisetyarso}(2011{\natexlab{a}})}]{trisetyarso2011erratum}
\bibinfo{author}{\bibfnamefont{A.}~\bibnamefont{Trisetyarso}},
  \bibinfo{journal}{Journal of Mathematical Physics}
  \textbf{\bibinfo{volume}{52}}, \bibinfo{pages}{019902}
  (\bibinfo{year}{2011}{\natexlab{a}}).

\bibitem[{\citenamefont{Trisetyarso and
  Silaban}(2002)}]{trisetyarso2008degeneracy}
\bibinfo{author}{\bibfnamefont{A.}~\bibnamefont{Trisetyarso}} \bibnamefont{and}
  \bibinfo{author}{\bibfnamefont{P.}~\bibnamefont{Silaban}}, Master's thesis,
  \bibinfo{school}{Institut Teknologi Bandung} (\bibinfo{year}{2002}).

\bibitem[{\citenamefont{Trisetyarso}(2013)}]{trisetyarso2013perturbation}
\bibinfo{author}{\bibfnamefont{A.}~\bibnamefont{Trisetyarso}}, in
  \emph{\bibinfo{booktitle}{Information and Communication Technology (ICoICT),
  2013 International Conference of}} (\bibinfo{organization}{IEEE},
  \bibinfo{year}{2013}), pp. \bibinfo{pages}{308--310}.

\bibitem[{\citenamefont{Trisetyarso
  et~al.}(2009{\natexlab{a}})\citenamefont{Trisetyarso, Van~Meter, and
  Itoh}}]{trisetyarso2009measurement}
\bibinfo{author}{\bibfnamefont{A.}~\bibnamefont{Trisetyarso}},
  \bibinfo{author}{\bibfnamefont{R.}~\bibnamefont{Van~Meter}},
  \bibnamefont{and} \bibinfo{author}{\bibfnamefont{K.~M.} \bibnamefont{Itoh}},
  in \emph{\bibinfo{booktitle}{8th Asian Conference on Quantum Information
  Science, KIAS, Seoul, Korea}} (\bibinfo{year}{2009}{\natexlab{a}}).

\bibitem[{\citenamefont{Trisetyarso
  et~al.}(2009{\natexlab{b}})\citenamefont{Trisetyarso, Van~Meter, and
  Itoh}}]{trisetyarso2009resources}
\bibinfo{author}{\bibfnamefont{A.}~\bibnamefont{Trisetyarso}},
  \bibinfo{author}{\bibfnamefont{R.}~\bibnamefont{Van~Meter}},
  \bibnamefont{and} \bibinfo{author}{\bibfnamefont{K.~M.} \bibnamefont{Itoh}},
  in \emph{\bibinfo{booktitle}{International Symposium on Nanoscale Transport
  and Technology 2009, NTT, Hon Atsugi, Japan}}
  (\bibinfo{year}{2009}{\natexlab{b}}).

\bibitem[{\citenamefont{Trisetyarso}(2011{\natexlab{b}})}]{trisetyarso2011theoretical}
\bibinfo{author}{\bibfnamefont{A.}~\bibnamefont{Trisetyarso}}, Ph.D. thesis,
  \bibinfo{school}{Keio University} (\bibinfo{year}{2011}{\natexlab{b}}).

\end{thebibliography}

\end{document}